Observation of excitonic Floquet states in a one-dimensional organic Mott insulator using mid-infrared pump near-infrared probe reflection spectroscopy

R. Ikeda[1*], D. Sakai[1], T. Yamakawa[1], T. Miyamoto[1,2], T. Hasegawa[3], H. Okamoto[1*]

[1] Department of Advanced Materials Science, University of Tokyo, Kashiwa 277-8561, Japan

[2] Department of Engineering, Nagoya Institute of Technology, Nagoya 466-8555, Japan

[3] Department of Applied Physics, University of Tokyo, Tokyo 113-8656, Japan.

*Correspondence author, email:

r.ikeda@edu.k.u-tokyo.ac.jp, okamotoh@k.u-tokyo.ac.jp

Abstract

When an electric field of light with a frequency of $\Omega$ is applied to a solid, Floquet states, consisting of sidebands with an interval of $\hbar\Omega$ around an electronic state, are expected to be formed. However, only a few studies have experimentally detected such sidebands. Here, we apply mid-infrared pump near-infrared reflection probe spectroscopy to a one-dimensional Mott insulator, bis(ethylendithio)tetrathianfulvalence-difluorotetracyanoquinodimethane (ET-F$_2$TCNQ), to detect the transient change in reflectivity $R$, $\Delta R/R$, due to the formation of excitonic Floquet states. Analyses, considering both odd- and even-parity excitons, demonstrate that the $\Delta R/R$ spectrum reflects the formation of the first-order Floquet sidebands of excitons, and its spectral shape strongly depends on the widths of excitonic states. The experimental and analytical approach reported here is effective in demonstrating excitonic Floquet states in various solids.

According to the Floquet's theorem, when a temporally periodic potential with frequency $\Omega$ is applied to a solid, sidebands at an interval of $\hbar\Omega$ are formed around an electronic state [1-3]. This quasi-stationary state, called a Floquet state, can be expressed as a superposition of wavefunctions of those sidebands. Floquet engineering attempts to create a new electronic state by generating such Floquet states [4–14].

Floquet states in solids can be generated by applying a strong electric field using a femtosecond laser pulse. Floquet states have been achieved by applying time-resolved angle-resolved photoemission spectroscopy (TR-ARPES) on weakly interacting electronic systems of topological materials and van der Waals semiconductors [15–19] having unique two-dimensional (2D) band structures. In one-dimensional (1D) systems having strong excitonic effects, sidebands are expected to be formed around excitonic states under light electric fields. In a 1D Mott insulator of a chlorine-bridged nickel-chain compound, subcycle spectroscopy was conducted using a phase-stable mid-infrared (MIR) pump pulse and an ultrashort visible probe pulse [20]. The changes in the intensity of the reflected probe light as well as large coherent oscillations, owing to quantum interferences between the reflected probe light and radiated lights from the sidebands, have been observed. The analyses of their spectra demonstrated the formation of Floquet sidebands. However, in subcycle spectroscopy, the available photon-energy range is limited [21,22]. Therefore, only a few observations of excitonic Floquet states have been reported.

In this study, we aimed to detect the sidebands of excitonic Floquet states using conventional MIR pump-near-IR (NIR) reflection probe spectroscopy. The target material was an organic compound, (bis(ethylendithio)tetrathianfulvalence-difluorotetracyano-quinodimethane (ET-$F_2$TCNQ) [23]. In ET-$F_2$TCNQ, an electron is transferred from donor molecules of ET to acceptor molecules of $F_2$TCNQ [Fig. 1(a)]. $F_2$TCNQ$^-$ molecules are nearly isolated, whereas ET molecules are stacked along the *a*-axis, forming a half-filled band [Fig. 1(a)]. Because of the large on-site Coulomb repulsion $U$ on ET overcoming the transfer integral $t$ between neighbouring ETs, electrons are localized on each ET, making this compound a 1D Mott insulator [24–26].

When this compound is irradiated with an MIR pulse, it is expected that the transition to a one-photon allowed exciton is suppressed and transitions to Floquet sidebands are increased. The pump-probe experiments detected large reflectivity changes around the original exciton peak. By the analysis incorporating third-order nonlinear optical responses of odd- and even-parity excitons, it was showed that the reflectivity changes originated from the formation of first-order Floquet sidebands.

In the MIR pump-NIR probe reflection spectroscopy, a Ti:sapphire regenerative amplifier (RA) with a central photon energy of 1.55 eV, temporal width of ~35 fs, pulse energy of 7 mJ, and repetition rate of 1 kHz was used as the light source. The RA output was divided into two beams with 6 and 1 mJ and were input to optical parametric amplifiers, OPA1 and OPA2, respectively. From OPA1 and OPA2, MIR pump pulses with 0.15 or 0.25 eV and NIR probe pulses from 0.50 to 1.1 eV were generated, respectively. The NIR pulses had temporal widths of ~55 fs. The reflected light was introduced into a 10-cm monochromator, and the central photon energy of each probe pulse was detected with a resolution of ~10 meV. All the measurements were performed at 60 K.

Figure 1(b) shows the spectra of reflectivity $R$ with the light electric field parallel to the *a*-axis and imaginary part of dielectric constant $\varepsilon_2$, obtained using the Kramers–Kronig (KK) transformation of the $R$ spectrum. The peak around 0.70 eV is due to the lowest one-photon allowed exciton (the doublon–holon bound state) with odd parity, $|\varphi_o\rangle$ [27,28]. Previously, electro-reflectance spectroscopy at 294 K using terahertz electric-field pulses showed that the one-photon forbidden exciton with even parity, $|\varphi_e\rangle$, was located at ~30 meV above $|\varphi_o\rangle$ and a doublon–holon continuum started at its higher energy side [Fig. 1(c)] [27,28].

The predicted energy-level structure of excitonic states under an MIR electric field is shown in Fig. 1(d) [20]. In 1D Mott insulators such as ET-F$_2$TCNQ, exciton wavefunctions of $|\varphi_o\rangle$ and $|\varphi_e\rangle$ have similar shapes except for their phases [Fig. 1(c)], and thus, the transition dipole moment between them, $\langle\varphi_o|x|\varphi_e\rangle$, becomes enhanced [27]. This causes the large third-order optical nonlinearity in 1D Mott insulators [29–33]. The transition dipole moment between $|\varphi_o\rangle$ and the ground state $|\varphi_g\rangle$, $\langle\varphi_g|x|\varphi_o\rangle$, is smaller than $\langle\varphi_o|x|\varphi_e\rangle$. Therefore, an MIR electric field generates only the sidebands associated with two excitons $|\varphi_o\rangle$ and $|\varphi_e\rangle$, represented by $|\varphi_o, \pm n\rangle$ and $|\varphi_e, \pm n\rangle$, respectively. Here, $n$ is a positive integer. The states $|\varphi_o, 0\rangle$ and $|\varphi_e, 0\rangle$ correspond to the original exciton states, $|\varphi_o\rangle$ and $|\varphi_e\rangle$, respectively [20]. In the Floquet states, optically allowed states are $|\varphi_o, 0\rangle$, $|\varphi_o, \pm 2n\rangle$, and $|\varphi_e, \pm(2n-1)\rangle$. Under the MIR electric field, it is expected that the intensity of the transition from $|\varphi_g\rangle$ to $|\varphi_o, 0\rangle$ reduces as compared with that of the original $|\varphi_g\rangle$ to $|\varphi_o\rangle$ transition, whereas the $|\varphi_g\rangle$ to $|\varphi_o, \pm 2n\rangle$ and $|\varphi_e, \pm(2n-1)\rangle$ transitions gain finite intensities [20]. We aimed to detect the changes in these transition intensities and thereby demonstrate the formations of Floquet sidebands.

In the MIR pump-NIR reflection probe spectroscopy [Fig. 2(a)], the pump photon energy was set to 0.15 or 0.25 eV. The spectra are presented in Fig. 2(b), together with the steady-state $R$ and $\varepsilon_2$ spectra, which show weak phonon absorptions. Since their intensities are weak, ET-F$_2$TCNQ can be considered fundamentally transparent to the pump lights. Figure 3(a) and (b) show the time characteristics of the reflectivity changes $\Delta R(t)/R$ at several probe energies induced by MIR pump pulses. The electric-field amplitudes are 0.66 and 0.45 MV/cm for the 0.15 and 0.25 eV pulses, respectively. The observed responses are symmetrical on the temporal axis, suggesting that they are coherent. The temporal width of $\Delta R(t)/R$ signals for the 0.25 eV excitation is ~120 fs, whereas that for the 0.15 eV excitation is broadened to ~450 fs. This is ascribed to the difference in the temporal widths of the MIR pulses: ~120 fs for the 0.25 eV pulse and ~450 fs for the 0.15 eV pulse. The broad width of the latter is because of the overlapping of its spectrum with the atmospheric water absorptions. The electric field amplitudes $E_{MIR}$ and temporal widths of MIR pulses are reported in Supplementary Material S1 [34].

The dependence of $\Delta R(0 \text{ ps})/R$ at 0.68 eV on the electric-field amplitude $E_{MIR}$ of the 0.15 eV pulse is shown in Fig. 3(c). $\Delta R(0 \text{ ps})/R$ is proportional to $[E_{MIR}]^2$ up to $E_{MIR} \sim 1.0$ MV/cm, indicating that the responses in this $E_{MIR}$ region are due to the third-order optical nonlinearity; the higher-order responses hardly appear. Here, $\Delta R(0 \text{ ps})/R$ is dominated by the transitions from $|\varphi_g\rangle$ to $|\varphi_o, 0\rangle$ and $|\varphi_e, \pm 1\rangle$.

The probe-energy dependences of $\Delta R(0 \text{ ps})/R$ for the 0.25-eV and 0.15-eV excitations are shown in Fig. 3(e) and (f), respectively, together with the $R$ spectrum [Fig. 3(d)]. Based on the result of terahertz electro-reflectance spectroscopy at 294 K, we assume that the energy of the even-parity exciton $|\varphi_e\rangle$, $\hbar\omega_e$, is 30 meV higher than that of the odd-parity exciton $|\varphi_o\rangle$, $\hbar\omega_o$. As $\hbar\omega_o$ is 0.68 eV at 60 K, the transition to $|\varphi_e, \pm 1\rangle$ is expected to appear at $\hbar\omega_e \pm \hbar\Omega$ ($\hbar\omega_e \sim 0.71$ eV), as indicated by the vertical bars in Fig. 3(e) and (f). The $\Delta R(0)/R$ spectrum covers the region from 0.71 eV $- \hbar\Omega$ to 0.71 eV $+ \hbar\Omega$ for $\hbar\Omega = 0.15$ eV but does not cover for $\hbar\Omega = 0.25$ eV. Hence, we analyzed the $\Delta R(0 \text{ ps})/R$ spectrum for the 0.15 eV excitation in detail.

The changes in the real and imaginary parts of the complex dielectric constant $\tilde{\varepsilon}(\omega) = \varepsilon_1(\omega) + i\varepsilon_2(\omega)$, $\Delta\varepsilon_1(\omega)$ and $\Delta\varepsilon_2(\omega)$, were obtained by the KK transform of $[R + \Delta R(0 \text{ ps})]$, shown in Fig. 3(g) and (h), respectively. $\Delta\varepsilon_2(\omega)$ represents positive, negative, and positive changes from the low energy side. These changes suggest that the $|\varphi_g\rangle$ to $|\varphi_o, 0\rangle$ transition becomes weaker than the original $|\varphi_g\rangle$ to $|\varphi_o\rangle$ transition, and $|\varphi_g\rangle$ to $|\varphi_e, \pm 1\rangle$ transitions appear. However, the absorption increases at ~0.60 and ~0.70 eV, which is inconsistent with the expected energies of $|\varphi_e, -1\rangle$ and $|\varphi_e, +1\rangle$, 0.71 eV $\pm \hbar\Omega$, that is, ~0.56 and ~0.86 eV, respectively.

To clarify this, a simulation was performed for the $\Delta\varepsilon_2(\omega)$ spectrum. First, the original $\varepsilon_2(\omega)$ spectrum was fitted by the following Lorentz oscillator:

$$\varepsilon_2(\omega) = \frac{Ne^2}{\hbar}\langle\varphi_g|x|\varphi_o\rangle^2 \text{Im}\left(\frac{1}{\omega_o - \omega - i\gamma_o} + \frac{1}{\omega_o + \omega + i\gamma_o}\right). \tag{1}$$

Here, $\hbar\omega_o$ and $\hbar\gamma_o$ are the energy and width of $|\varphi_o\rangle$, respectively, and $\langle\varphi_g|x|\varphi_o\rangle$ is the transition dipole moment between $|\varphi_g\rangle$ and $|\varphi_o\rangle$. The fitting curve [the red broken line in Fig. 1(b)] with the parameters in Table 1 reproduced the experimental results well.

Then, we analyzed the $\Delta\varepsilon_1(\omega)$ and $\Delta\varepsilon_2(\omega)$ spectra using Eq. (S1) of $\chi^{(3)}(-\omega; \Omega, -\Omega, \omega)$ (Supplemental Material S2 [34]) based on the three-level model, in which $|\varphi_g\rangle$, $|\varphi_o\rangle$, and $|\varphi_e\rangle$ are considered [29,35]. In the framework of the third-order optical nonlinearity, the following relationship stands between $\Delta\tilde{\varepsilon}(\omega)$ and $\chi^{(3)}(-\omega; \Omega, -\Omega, \omega)$ [35]:

$$\Delta\tilde{\varepsilon}(\omega) = \frac{3}{2}\chi^{(3)}(-\omega; \Omega, -\Omega, \omega)E(\Omega)E(-\Omega) \tag{2}$$

As $\hbar\omega_o$, $\hbar\gamma_o$, and $\langle\varphi_g|x|\varphi_o\rangle$ had been determined (Table 1), the fitting parameters were the energy, $\hbar\omega_2$, and width, $\hbar\gamma_2$, of $|\varphi_e\rangle$, and the transition dipole moment $\langle\varphi_o|x|\varphi_e\rangle$ between $|\varphi_o\rangle$ and $|\varphi_e\rangle$. However, the $\Delta\varepsilon_1(\omega)$ and $\Delta\varepsilon_2(\omega)$ spectra could not be reproduced.

Therefore, we next attempted to reproduce the $\Delta\varepsilon_1(\omega)$ and $\Delta\varepsilon_2(\omega)$ spectra by the four-level model, in which another one-photon allowed state with odd parity, $|\varphi_{o2}\rangle$, on the higher energy side of $|\varphi_e\rangle$ was added to the three-level model [27,32,36] (Supplementary Material S2 [34]). This four-level model could explain the $\Delta\varepsilon_2(\omega)$ and $\chi^{(3)}(-\omega; 0, 0, \omega)$ spectra in ET-F$_2$TCNQ obtained by the terahertz electro-reflectance spectroscopy [27]. In this model, the energy and width of $|\varphi_{o2}\rangle$ ($\hbar\omega_{o2}$ and $\hbar\gamma_{o2}$, respectively), and the transition dipole moments $\langle\varphi_e|x|\varphi_{o2}\rangle$ and $\langle\varphi_g|x|\varphi_{o2}\rangle$ are additional parameters. Because $\langle\varphi_e|x|\varphi_{o2}\rangle$ and $\langle\varphi_g|x|\varphi_{o2}\rangle$ appear as their product in the formula of $\chi^{(3)}(-\omega; \Omega, -\Omega, \omega)$, they cannot be determined independently and $\langle\varphi_g|x|\varphi_{o2}\rangle\langle\varphi_{o2}|x|\varphi_e\rangle$ is a fitting parameter.

Using the four-level model, the $\Delta\varepsilon_1(\omega)$, $\Delta\varepsilon_2(\omega)$, and $\Delta R(0\text{ ps})/R$ spectra were well reproduced [red lines in Fig. 3(f–h)] with the parameter values listed in Table 1. In Fig. 3(g) and (h), the energy of $|\varphi_o\rangle$, $\hbar\omega_1$, and the predicted energies of $|\varphi_e, \pm 1\rangle$, $\hbar\omega_2 \pm 0.15$ eV, are indicated by green and blue arrows, respectively. The dip around 0.68 eV in the $\Delta\varepsilon_2(\omega)$ spectrum is owing to the reduction in the intensity of the $|\varphi_g\rangle \to |\varphi_o, 0\rangle$ transition. The calculated $\Delta\varepsilon_2(\omega)$ spectrum shows that the absorption increases around 0.60 and 0.70 eV, which are consistent with the experimental results but differ from the energy positions of $|\varphi_e \pm 1\rangle$, predicted from Fig. 1(d).

A possible reason is that the widths of exciton states cannot be neglected compared to the energy interval between $|\varphi_o\rangle$ and $|\varphi_e, +1\rangle$ or $|\varphi_e, -1\rangle$, which is $(\hbar\omega_e + \hbar\Omega) - \hbar\omega_o \sim 0.20$ eV or $\hbar\omega_o - (\hbar\omega_e - \hbar\Omega) \sim 0.10$ eV. The width of $|\varphi_o\rangle$, $\hbar\gamma_o$, is 0.0152 eV, which is sufficiently small compared to 0.15 eV, whereas the width of $|\varphi_e\rangle$, $\hbar\gamma_e = 0.050$ eV, is comparable (Table 1). Therefore, it is natural to assume that $\hbar\gamma_e$ has a significant influence on the spectral shapes of $\Delta\varepsilon_1(\omega)$ and $\Delta\varepsilon_2(\omega)$.

To further clarify this, we calculated the $\hbar\gamma_e$ dependence of $\Delta\varepsilon_2(\omega)$ with the simpler three-level model using the parameters $\hbar\omega_{o,e}$, $\hbar\gamma_{o,e}$, $\langle\varphi_g|x|\varphi_o\rangle$, and $\langle\varphi_o|x|\varphi_e\rangle$ in Table 1. In the $\Delta\varepsilon_2(\omega)$ spectrum for $\hbar\gamma_e = 0.020$ eV [Fig. 4(a)], the absorption increases at the energies $\hbar\omega_e \pm \hbar\Omega$ (the broken lines) predicted for $|\varphi_e, \pm 1\rangle$. The decrease in absorption near the original $\varepsilon_2(\omega)$ peak is asymmetric; the negative component of $\Delta\varepsilon_2(\omega)$ is enhanced at the lower energy side and a small positive peak appears at the high energy side. Thus, the $|\varphi_g\rangle$ to $|\varphi_o, 0\rangle$ transition is not only reduced but also shifted to the higher energy [upper part of Fig. 4(b)]. As the energy of $|\varphi_o\rangle$ is lower than that of $|\varphi_e\rangle$, $|\varphi_e, -1\rangle$ is closer to $|\varphi_o, 0\rangle$ compared with $|\varphi_e, +1\rangle$, and $|\varphi_e, -1\rangle$ interacts with $|\varphi_o, 0\rangle$ more strongly than $|\varphi_e, +1\rangle$. This causes the state $|\varphi_o, 0\rangle$ to be pushed to a higher energy than $|\varphi_o\rangle$.

Figure 4(c) shows the $\Delta\varepsilon_2(\omega)$ spectra when $\hbar\gamma_e$ is varied from 0.02 to 0.10 eV with the same values of the other parameters. As $\hbar\gamma_e$ increases, the peaks due to $|\varphi_e, +1\rangle$ and $|\varphi_e, -1\rangle$ become broad. Simultaneously, a positive absorption change appears on the slightly lower energy side of the absorption decrease at 0.68 eV, and the positive absorption change, observed for $\hbar\gamma_e = 0.020$ eV on the higher energy side, is enhanced. This can be explained as follows. Absorptions due to $|\varphi_e, \pm 1\rangle$ should appear at the energies where the density of state of $|\varphi_e, \pm 1\rangle$ is high. If $\hbar\gamma_e$ is sufficiently smaller than $\hbar\Omega = 0.15$ eV, this factor dominates the induced absorptions and the absorption peak due to $|\varphi_e, +1\rangle$ ($|\varphi_e, -1\rangle$) appears at $\hbar\omega_e + \hbar\Omega$ ($\hbar\omega_e - \hbar\Omega$) [blue triangles in Fig. 4(b)]. As $\hbar\gamma_e$ increases, the overlap between the states of $|\varphi_e, +1\rangle$ ($|\varphi_e, -1\rangle$) and $|\varphi_o\rangle$ increases [upper part of Fig. 4(d)], which strengthens the effective interaction between them. This enhances the absorptions in the region close to $|\varphi_o\rangle$. When this effect becomes dominant, the absorption at the position of $|\varphi_e, +1\rangle$ ($|\varphi_e, -1\rangle$) is relatively decreased, but that at the energies somewhat closer to $|\varphi_o\rangle$, increases. In fact, for $\hbar\gamma_e = 0.08$ eV, new structures appear around 0.64 and 0.74 eV [green triangles in Fig. 4(d)], leaving weak peak structures around 0.58 and 0.85 eV, corresponding to $|\varphi_e, -1\rangle$ and $|\varphi_e, +1\rangle$, respectively [blue triangles in Fig. 4(d)]. In the experiments, the peak structures were observed around 0.64 and 0.58 eV on the low energy side [Fig. 3(h)]. On the high energy side, the features observed in Fig. 4(d) were not as clearly seen, owing to the presence of $|\varphi_{o2}\rangle$. As the oscillator strengths of induced absorptions were moved from the $|\varphi_g\rangle$ to $|\varphi_o\rangle$ transition, a

sharp absorption decrease appeared around $\hbar\omega_{\mathrm{o}}$, regardless of $\hbar\gamma_e$ values. Thus, the states $|\varphi_{\mathrm{e}}, +1\rangle$, $|\varphi_{\mathrm{o}}, 0\rangle$, and $|\varphi_{\mathrm{e}}, -1\rangle$ were identified and the formation of excitonic Floquet states was demonstrated.

For the 0.25-eV excitation, $\Delta\tilde{\varepsilon}(\omega)$ could not be obtained because of the lack of low-energy $\Delta R/R$ spectrum. Therefore, we directly reproduced the $\Delta R/R$ spectrum by the four-level model using the parameters in Table 1. The calculated spectrum [red line in Fig. 3(e)], multiplied by 4.2, nearly reproduced the $\Delta R/R$ spectrum. The difference in magnitude of the experimental and calculated spectra might be attributed to an error in the estimation of $E_{\mathrm{MIR}}$. Thus, the analytical method using the four-level model is effective to calculate nonlinear optical spectra associated with the Floquet state in 1D Mott insulators.

In summary, the transient reflectivity spectrum induced by the MIR electric field was measured in a 1D Mott insulator, ET-F$_2$TCNQ. By analyzing the spectral changes of the complex dielectric constants derived from the transient reflectivity spectrum, the first-order Floquet sidebands were identified and the formation of the excitonic Floquet states was demonstrated. The analysis also revealed that the spectral changes originating from the Floquet sidebands strongly depend on the width of excitonic states. Our approach is effective because Floquet states can be observed using conventional pump-probe spectroscopy. Here, only the first-order sidebands corresponding to the third-order optical nonlinearity were observed. It is expected that similar experiments using an MIR pulse with a larger electric-field amplitude can detect sidebands of the second order and higher.

**Acknowledgements**

We thank Dr. Y. Takahashi (Hokkaido Univ.) for his support of the sample preparations in the early stage of this work. This work was supported in part by Grants-in-Aid for Scientific

Research from the Japan Society for the Promotion of Science (JSPS) (Project Numbers: JP21H04988, JP20K03801, JP18H01166), Grant-in-Aid for JSPS Fellows (Project Number: JP21J22162) and by CREST (Grant Number: JPMJCR1661), Japan Science and Technology Agency. T.Y. was supported by the Program for Leading Graduate Schools (MERIT-WINGS) and JSPS Research Fellowship for Young Scientists. D.S. was supported by Support for Pioneering Research Initiated by Next Generation of Japan Science and Technology Agency (JST SPRING) (Grant Number: JPMJSP2108).

interaction in two-dimensional cuprate Mott insulators", Science Advances **5**, eaav2187 (2019).

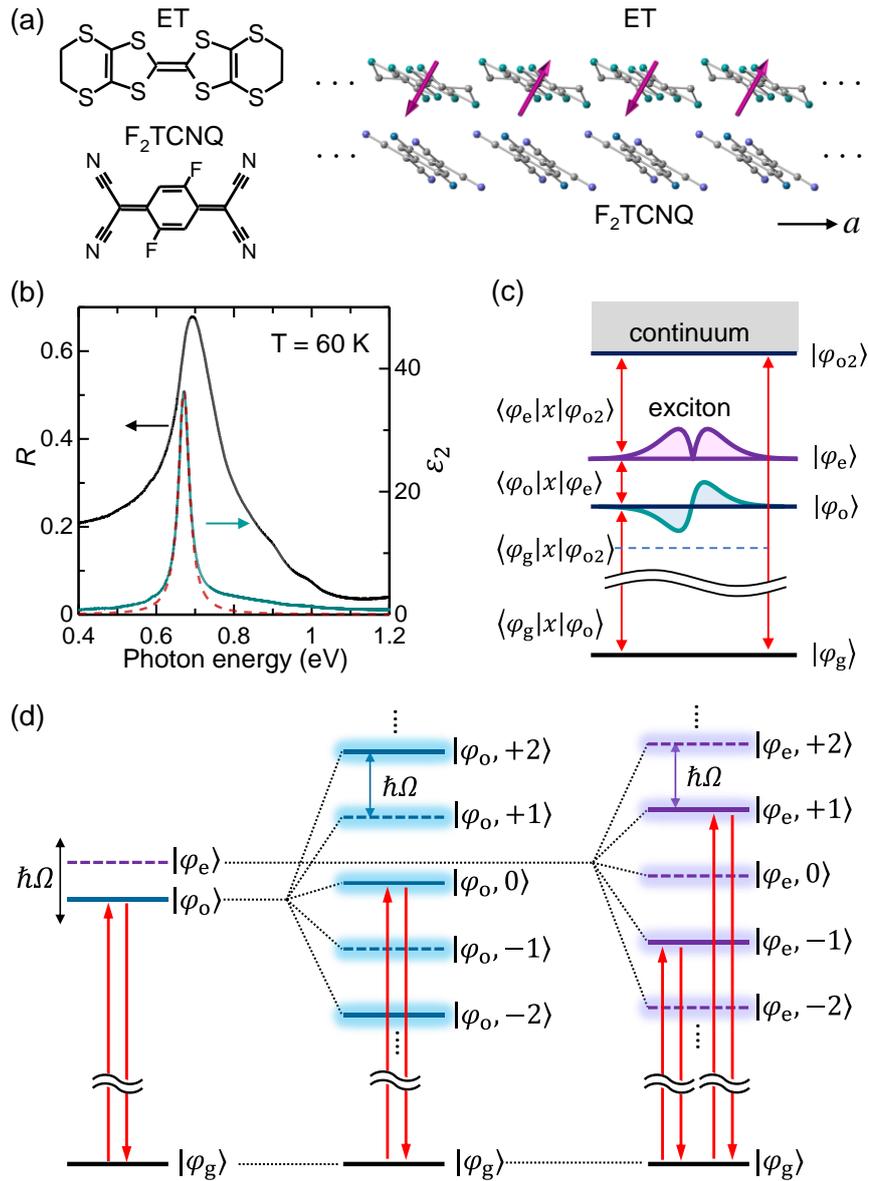

Fig. 1 (a) Molecular and crystal structures of ET-F$_2$TCNQ; (b) $R$ and $\varepsilon_2$ spectra along the $a$-axis; (c) Conceptual diagram of excited states in ET-F$_2$TCNQ; (d) Conceptual diagram of Floquet sidebands associated with odd- and even-parity excitons under the presence of an MIR electric field with frequency of $\hbar\Omega$. Solid and dotted lines: one-photon allowed and forbidden states, respectively; blue arrows: transitions related to third-order nonlinear optical responses

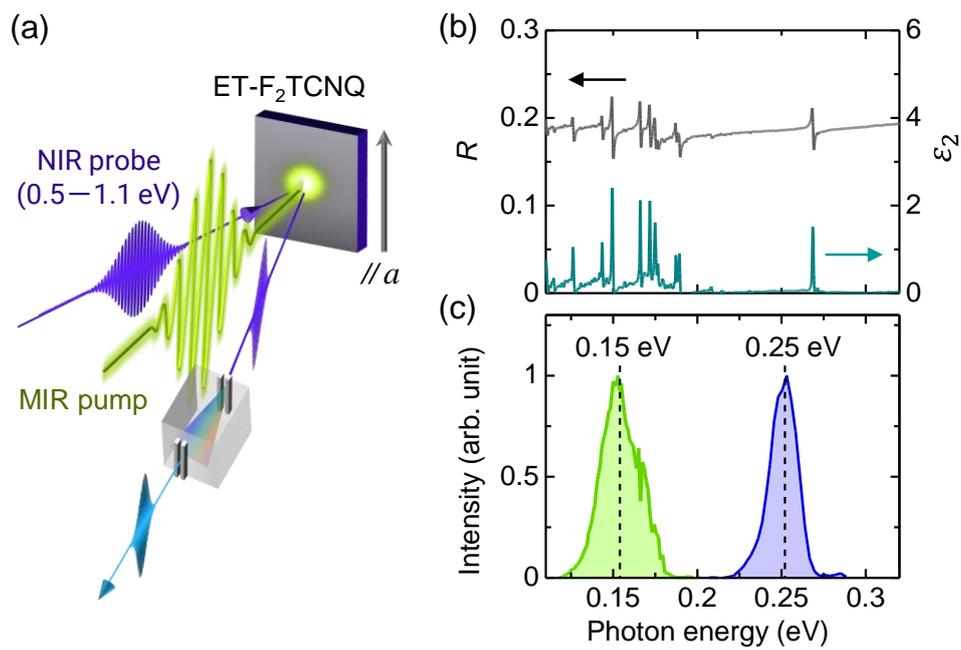

Fig. 2 (a) Schematic of MIR pump-NIR probe reflection spectroscopy; (b) $R$ spectrum along the $a$-axis; (c) Spectra of MIR pump pulses in this study

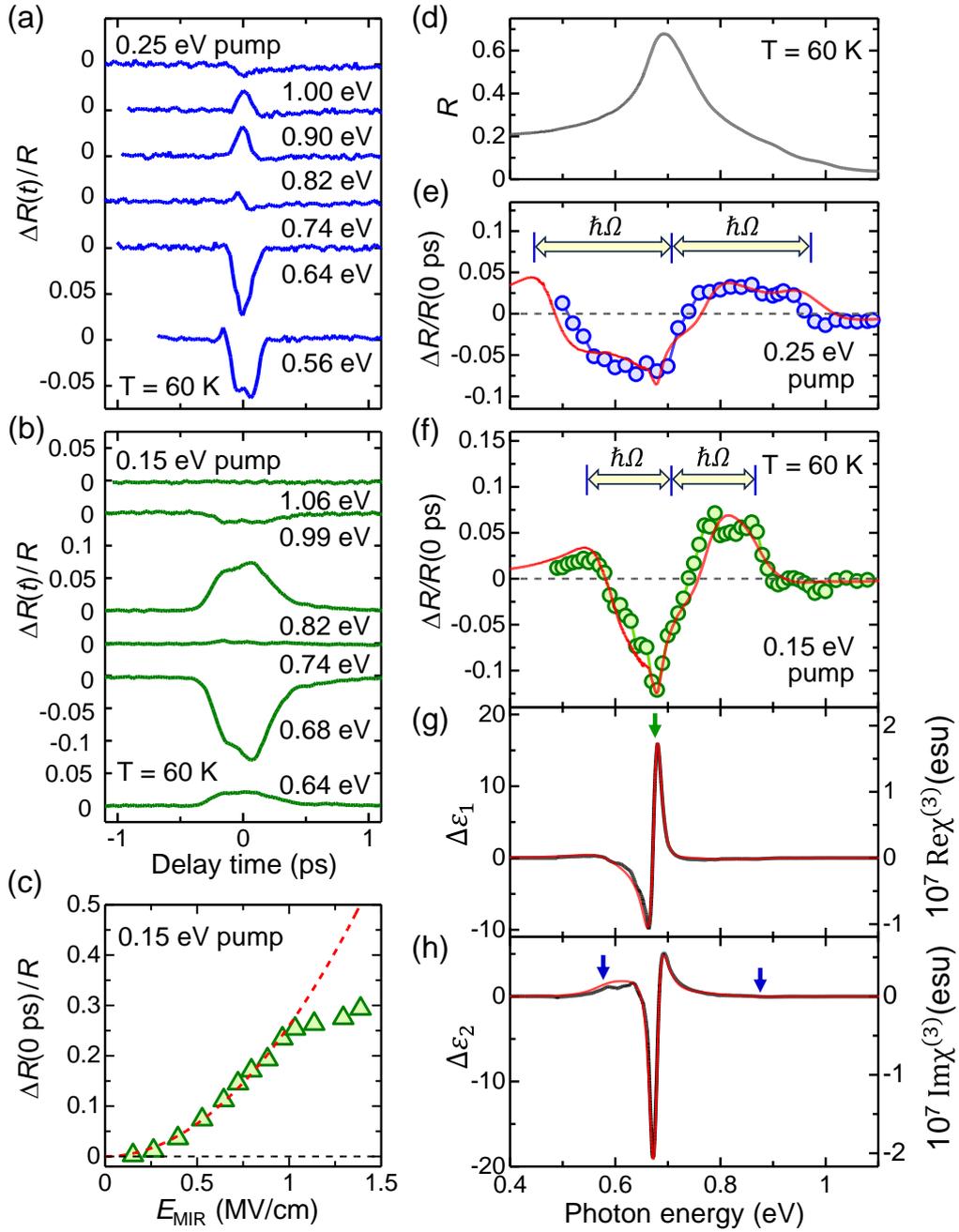

Fig. 3 (a,b) Time characteristics of the reflectivity change $\Delta R(t)/R$ by the MIR pulse: (a) $\hbar\Omega = 0.25$ eV, (b) $\hbar\Omega = 0.15$ eV; (c) Dependence of $-\Delta R(0\text{ ps})/R$ at 0.68 eV on the electric field amplitude $E_{\text{MIR}}$ for $\hbar\Omega = 0.15$ eV. Broken line: relation $-\Delta R(0\text{ ps})/R \propto E_{\text{MIR}}$. (d) $R$ spectrum along the $a$-axis (same as spectrum in Fig. 1(b)); (e,f) Probe energy dependence of $\Delta R(0\text{ ps})/R$: (e) $\hbar\Omega = 0.25$ eV, (f) $\hbar\Omega = 0.15$ eV; (g, h) $\Delta\varepsilon_1(\omega)$ and $\Delta\varepsilon_2(\omega)$ spectra derived from $\Delta R(0\text{ ps})/R$ spectrum for $\hbar\Omega = 0.15$ eV. Red lines in (e–h): fitting curves (see text)

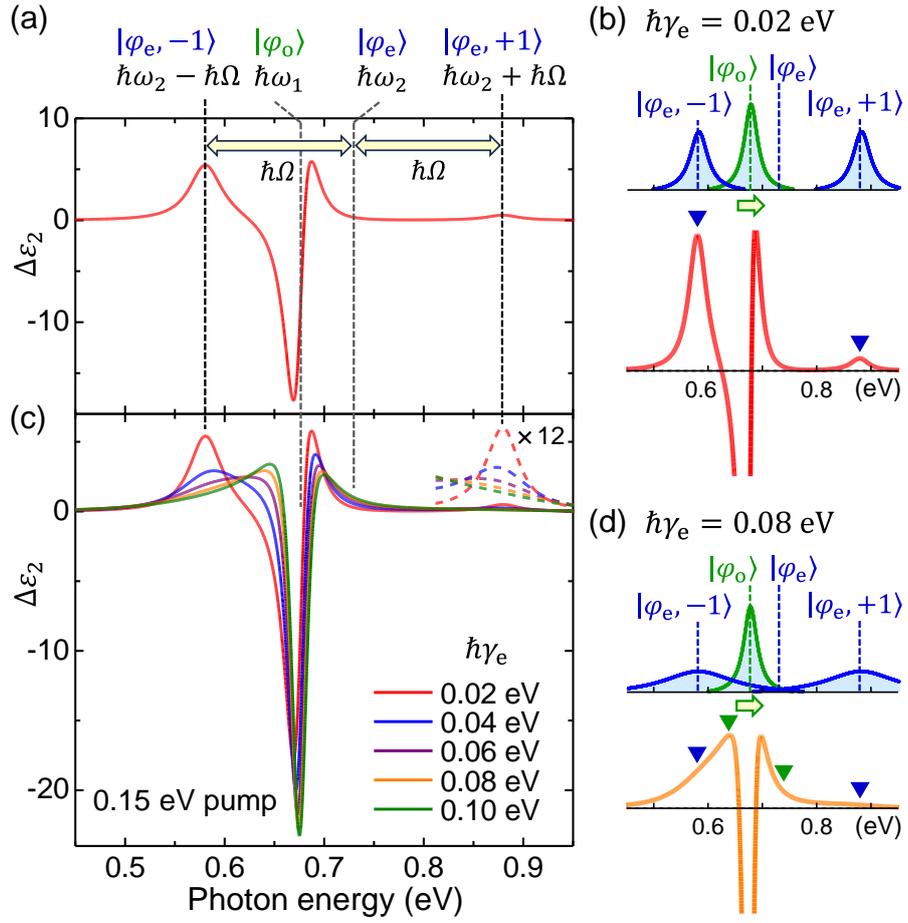

Fig. 4 (a) Simulation of $\Delta\varepsilon_2$ spectrum based upon three-level model. Pump photon energy: 0.15 eV; width of even-parity exciton, $\hbar\gamma_2$: 0.02 eV. (b) Schematic of $\Delta\varepsilon_2$ spectrum when $\hbar\gamma_2 = 0.02$ eV; (c) Simulations of $\Delta\varepsilon_2$ spectra based upon three-level model for various $\hbar\gamma_2$ values, 0.02–0.10 eV. Pump photon energy: 0.15 eV. (d) Schematic of $\Delta\varepsilon_2$ spectrum when $\hbar\gamma_2 = 0.08$ eV

Table 1. Fitting parameters of $\varepsilon_2(\omega)$ in Eq. (1) and $\Delta\varepsilon_2(\omega)$ in the four-level model. The parameters for the fittings of $\varepsilon_2(\omega)$ and $\Delta\varepsilon_2(\omega)$ are accurate to three and two significant digits, respectively.

| $\hbar\omega_{o1}$ (eV) | $\hbar\gamma_{o1}$ (eV) | $\hbar\omega_e$ (eV) | $\hbar\gamma_e$ (eV) | $\hbar\omega_{o2}$ (eV) | $\hbar\gamma_{o2}$ (eV) |
|---|---|---|---|---|---|
| 0.676 | 0.0152 | 0.73 | 0.050 | 0.78 | 0.080 |

| $\langle\varphi_g|x|\varphi_{o1}\rangle$ (Å) | $\langle\varphi_o|x|\varphi_e\rangle$ (Å) | $\langle\varphi_g|x|\varphi_{o2}\rangle\langle\varphi_{o2}|x|\varphi_e\rangle$ (Å$^2$) |
|---|---|---|
| 1.29 | 9.8 | -5.3 |

**Supplemental Materials:**

Observation of excitonic Floquet states in a one-dimensional organic Mott insulator using mid-infrared pump near-infrared probe reflection spectroscopy

R. Ikeda[1*], D. Sakai[1], T. Yamakawa[1], T. Miyamoto[1,2], T. Hasegawa[3], H. Okamoto[1*]

[1] Department of Advanced Materials Science, University of Tokyo, Kashiwa 277-8561, Japan

[2] Department of Engineering, Nagoya Institute of Technology, Nagoya 466-8555, Japan

[3] Department of Applied Physics, University of Tokyo, Tokyo 113-8656, Japan.

*Correspondence author,

email: r.ikeda@edu.k.u-tokyo.ac.jp, okamotoh@k.u-tokyo.ac.jp

**Supplementary Material S1. Estimations of electric fields of MIR pulses**

The maximum electric field intensity $E_{\mathrm{MIR}}$ of an MIR pump pulse can be estimated from its temporal width and its fluence and spot size at the sample surface. The temporal width of the MIR pump pulse $\tau_{\mathrm{p}}$ is calculated from the temporal width of the probe pulse $\tau_{\mathrm{r}}$ and the temporal characteristic of the reflectivity change due to the coherent response shown in Fig. 2(a) and (b).

$\tau_{\mathrm{r}}$ is calculated assuming the Fourier limit and using the spectral width (full width at half maximum: FWHM) of the probe pulse. The spectral width (FWHM) of the probe pulse in the frequency domain is approximately 8 THz independent of the probe frequency in common, from which $\tau_{\mathrm{r}}$ is determined to be 55 fs. As observed in Fig. 2(a) and (b), the temporal width of the reflectivity change due to the coherent response is ~135 fs for the 0.25 eV pulse and ~450 fs for the 0.15 eV pulse. Here, it is assumed that the temporal profiles of the pump and probe pulses are expressed by Gaussian functions $I_{\mathrm{p}}(t) = \exp(-t^2/\tau_{\mathrm{p}}^2)$ and $I_{\mathrm{r}}(t) = \exp(-t^2/\tau_{\mathrm{r}}^2)$, respectively, and that the time characteristic of the coherent response in the reflectivity change $\Delta R(t)/R$ is also expressed by a Gaussian function as $\Delta R(t)/R \propto \exp(-t^2/\tau_{\mathrm{c}})$ with $\tau_{\mathrm{c}}$.

Because $\tau_c$ characterizing the coherent response is determined by the convolution integral of the pump and probe pulses, a relation $\tau_c^2 = \tau_p^2 + \tau_r^2$ stands between $\tau_p$, $\tau_r$, and $\tau_c$. Using this relation, the temporal widths of the 0.25-eV and 0.15-eV pulses are evaluated to be ~123 and ~446 fs, respectively. From the spot size of 100 μm (96 μm) and the temporal width of 123 fs (446 fs), the electric-field amplitude of the MIR pump pulse with 0.25 eV (0.15 eV) used for the measurements shown in Fig. 2(a) [Fig. 2(b)] is calculated to be 0.66 MV/cm (0.45 MV/cm) from the pulse fluence.

**Supplementary Material S2. Analyses of spectral changes under the MIR electric field based upon third-order optical nonlinearity**

The third-order nonlinear susceptibility $\chi^{(3)}(-\omega_\sigma; \omega_i, \omega_j, \omega_k)$ is defined by the following formula:

$$P^{(3)}(\omega_\sigma) = 2^{s_\omega - 2} p_\omega \chi^{(3)}(-\omega_\sigma; \omega_i, \omega_j, \omega_k) E(\omega_i) E(\omega_j) E(\omega_k) \tag{S1}$$

Here, $P^{(3)}(\omega_\sigma)$ is the third-order nonlinear polarization at frequency $\omega_\sigma$, $E(\omega_i)$ is the electric field of the incident pulses at frequency $\omega_i$, $s_\omega$ is the number of zeros among $(\omega_i, \omega_j, \omega_k)$, $p_\omega$ is the number of permutations of $(\omega_i, \omega_j, \omega_k)$, where $\omega_\sigma = \omega_i + \omega_j + \omega_k$. For the experimental condition of the MIR pump-NIR reflection probe spectroscopy, the change in the complex dielectric constant $\tilde{\varepsilon}(\omega)$, $\Delta\tilde{\varepsilon}(\omega)$, through the third-order nonlinear process is expressed as follows:

$$\Delta\tilde{\varepsilon}(\omega) = \frac{3}{2} \chi^{(3)}(-\omega; \Omega, -\Omega, \omega) E(\Omega) E(-\Omega) \tag{S2}$$

Here, $\Omega$ and $\omega$ are frequencies of the MIR pump pulse and NIR probe pulse, respectively.

The expression for $\chi^{(3)}(-\omega_\sigma; \omega_i, \omega_j, \omega_k)$ derived from perturbation theory is given by the following formula [35]:

$$\chi^{(3)}(-\omega_\sigma; \omega_i, \omega_j, \omega_k)$$
$$= \frac{Ne^4}{6\varepsilon_0 \hbar^3} \wp \sum_{abc} \langle \varphi_g | x | \varphi_a \rangle \langle \varphi_a | x | \varphi_b \rangle \langle \varphi_b | x | \varphi_c \rangle \langle \varphi_c | x | \varphi_g \rangle$$

$$\times \left[ \frac{1}{(\omega_a - \omega_\sigma - i\gamma_a)(\omega_b - \omega_j - \omega_k - i\gamma_b)(\omega_c - \omega_k - i\gamma_c)} \right.$$

$$+ \frac{1}{(\omega_a + \omega_i + i\gamma_a)(\omega_b - \omega_j - \omega_k - i\gamma_b)(\omega_c - \omega_k - i\gamma_c)}$$

$$+ \frac{1}{(\omega_a + \omega_i + i\gamma_a)(\omega_b + \omega_i + \omega_j + i\gamma_b)(\omega_c - \omega_k - i\gamma_c)}$$

$$\left. + \frac{1}{(\omega_a + \omega_i + i\gamma_a)(\omega_b + \omega_i + \omega_j + i\gamma_b)(\omega_c + \omega_\sigma + i\gamma_c)} \right] \quad (S3)$$

$N$ denotes the density of the ET molecules, $\varepsilon_0$ is the permittivity of the vacuum, $e$ is the elementary charge, $\hbar$ is the reduced Plank constant, $|\varphi_g\rangle$ is the wavefunction of the ground state, $|\varphi_a\rangle, |\varphi_b\rangle$ and $|\varphi_c\rangle$ are those of the excited states, $\langle\varphi_l|x|\varphi_m\rangle$ is the transition dipole moment between $|\varphi_l\rangle$ and $|\varphi_m\rangle$, $\hbar\omega_l$ and $\hbar\gamma_l$ are the energy and width of $|\varphi_l\rangle$, respectively, and $\wp$ denotes the permutation of $(\omega_i, \omega_j, \omega_k)$.

Assuming the three-level model with the ground state $|\varphi_g\rangle$, the odd-parity exciton $|\varphi_o\rangle$, and the even-parity exciton $|\varphi_e\rangle$, $\chi^{(3)}(-\omega_\sigma; \omega_i, \omega_j, \omega_k)$ is given as follows [35]:

$$\chi^{(3)}(-\omega_\sigma; \omega_i, \omega_j, \omega_k)$$

$$= \frac{Ne^4}{6\varepsilon_0 \hbar^3} \langle\varphi_g|x|\varphi_o\rangle\langle\varphi_o|x|\varphi_e\rangle\langle\varphi_e|x|\varphi_o\rangle\langle\varphi_o|x|\varphi_g\rangle$$

$$\times \wp \left[ \frac{1}{(\omega_o - \omega_\sigma - i\gamma_o)(\omega_e - \omega_j - \omega_k - i\gamma_e)(\omega_o - \omega_k - i\gamma_o)} \right.$$

$$+ \frac{1}{(\omega_o + \omega_i + i\gamma_o)(\omega_e - \omega_j - \omega_k - i\gamma_e)(\omega_o - \omega_k - i\gamma_o)}$$

$$+ \frac{1}{(\omega_o + \omega_i + i\gamma_o)(\omega_e + \omega_i + \omega_j + i\gamma_e)(\omega_o - \omega_k - i\gamma_o)}$$

$$\left. + \frac{1}{(\omega_o + \omega_i + i\gamma_o)(\omega_e + \omega_i + \omega_j + i\gamma_e)(\omega_o + \omega_\sigma + i\gamma_o)} \right] \quad (S4)$$

Because $\langle\varphi_o|x|\varphi_e\rangle\langle\varphi_e|x|\varphi_o\rangle \gg \langle\varphi_g|x|\varphi_o\rangle\langle\varphi_o|x|\varphi_g\rangle$ in 1D Mott insulators, the terms proportional to $\langle\varphi_g|x|\varphi_o\rangle\langle\varphi_o|x|\varphi_g\rangle\langle\varphi_g|x|\varphi_o\rangle\langle\varphi_o|x|\varphi_g\rangle$ are neglected. For the experimental condition of MIR pump-NIR reflection probe spectroscopy, $\chi^{(3)}(-\omega; \Omega, -\Omega, \omega)$ is given by the following 24 terms [35]:

$$\chi^{(3)}(-\omega;\Omega,-\Omega,\omega)$$

$$= \frac{Ne^4}{6\varepsilon_0 \hbar^3} \langle \varphi_g|x|\varphi_o\rangle\langle \varphi_o|x|\varphi_e\rangle\langle \varphi_e|x|\varphi_o\rangle\langle \varphi_o|x|\varphi_g\rangle$$

$$\times \left[ \frac{1}{(\omega_o - i\gamma_o - \omega)^2(\omega_e - i\gamma_e + \Omega - \omega)} + \frac{1}{(\omega_o - i\gamma_o - \omega)(\omega_e - i\gamma_e + \Omega - \omega)(\omega_o - i\gamma_o + \Omega)} \right.$$

$$+ \frac{1}{(\omega_o - i\gamma_o - \omega)^2(\omega_e - i\gamma_e - \Omega - \omega)} + \frac{1}{(\omega_o - i\gamma_o - \omega)(\omega_e - i\gamma_e - \Omega - \omega)(\omega_o - i\gamma_o - \Omega)}$$

$$+ \frac{1}{(\omega_o - i\gamma_o - \omega)(\omega_e - i\gamma_e)(\omega_o - i\gamma_o + \Omega)} + \frac{1}{(\omega_o - i\gamma_o - \omega)(\omega_e - i\gamma_e)(\omega_o - i\gamma_o - \Omega)}$$

$$+ \frac{1}{(\omega_o + i\gamma_o + \Omega)(\omega_e - i\gamma_e + \Omega - \omega)(\omega_o - i\gamma_o - \omega)} + \frac{1}{(\omega_o + i\gamma_o + \Omega)(\omega_e - i\gamma_e + \Omega - \omega)(\omega_o - i\gamma_o + \Omega)}$$

$$+ \frac{1}{(\omega_o + i\gamma_o - \Omega)(\omega_e - i\gamma_e - \Omega - \omega)(\omega_o - i\gamma_o - \omega)} + \frac{1}{(\omega_o + i\gamma_o - \Omega)(\omega_e - i\gamma_e - \Omega - \omega)(\omega_o - i\gamma_o - \Omega)}$$

$$+ \frac{1}{(\omega_o + i\gamma_o + \omega)(\omega_e - i\gamma_e)(\omega_o - i\gamma_o + \Omega)} + \frac{1}{(\omega_o + i\gamma_o + \omega)(\omega_e - i\gamma_e)(\omega_o - i\gamma_o - \Omega)}$$

$$+ \frac{1}{(\omega_o + i\gamma_o + \Omega)(\omega_e + i\gamma_e)(\omega_o - i\gamma_o - \omega)} + \frac{1}{(\omega_o + i\gamma_o + \Omega)(\omega_e + i\gamma_e + \Omega + \omega)(\omega_o - i\gamma_o + \Omega)}$$

$$+ \frac{1}{(\omega_o + i\gamma_o - \Omega)(\omega_e + i\gamma_e)(\omega_o - i\gamma_o - \omega)} + \frac{1}{(\omega_o + i\gamma_o - \Omega)(\omega_e + i\gamma_e - \Omega + \omega)(\omega_o - i\gamma_o - \Omega)}$$

$$+ \frac{1}{(\omega_o + i\gamma_o + \omega)(\omega_e + i\gamma_e + \Omega + \omega)(\omega_o - i\gamma_o + \Omega)}$$

$$+ \frac{1}{(\omega_o + i\gamma_o + \omega)(\omega_e + i\gamma_e - \Omega + \omega)(\omega_o - i\gamma_o - \Omega)}$$

$$+ \frac{1}{(\omega_o + i\gamma_o + \Omega)(\omega_e + i\gamma_e)(\omega_o + i\gamma_o + \omega)} + \frac{1}{(\omega_o + i\gamma_o + \Omega)(\omega_e + i\gamma_e + \Omega + \omega)(\omega_o + i\gamma_o + \omega)}$$

$$+ \frac{1}{(\omega_o + i\gamma_o - \Omega)(\omega_e + i\gamma_e)(\omega_o + i\gamma_o + \omega)} + \frac{1}{(\omega_o + i\gamma_o - \Omega)(\omega_e + i\gamma_e - \Omega + \omega)(\omega_o + i\gamma_o + \omega)}$$

$$\left. + \frac{1}{(\omega_o + i\gamma_o + \omega)^2(\omega_e + i\gamma_e + \Omega + \omega)} + \frac{1}{(\omega_o + i\gamma_o + \omega)^2(\omega_e + i\gamma_e - \Omega + \omega)} \right]. \quad (S5)$$

The dominant terms in Eq. (S5) are the first and third ones. However, the $\Delta\tilde{\varepsilon}(\omega)$ spectra experimentally obtained [black lines in Fig. 3(g,h)] cannot be reproduced by the results of the fitting analysis using this formula (not shown).

Next, we consider the four-level model, in which another odd-parity excited state, $|\varphi_{o2}\rangle$, is added to the three-level model on the higher energy side than $|\varphi_e\rangle$. In this model, $\chi^{(3)}(-\omega_\sigma;\omega_i,\omega_j,\omega_k)$ is given as follows [35]:

$$\chi^{(3)}(-\omega_\sigma;\omega_i,\omega_j,\omega_k)$$

$$= \frac{Ne^4}{6\varepsilon_0\hbar^3} \langle\varphi_g|x|\varphi_o\rangle\langle\varphi_o|x|\varphi_e\rangle\langle\varphi_e|x|\varphi_o\rangle\langle\varphi_o|x|\varphi_g\rangle$$

$$\times \wp \left[ \frac{1}{(\omega_o - \omega_\sigma - i\gamma_o)(\omega_e - \omega_j - \omega_k - i\gamma_e)(\omega_o - \omega_k - i\gamma_o)} \right.$$

$$+ \frac{1}{(\omega_o + \omega_i + i\gamma_o)(\omega_e - \omega_j - \omega_k - i\gamma_e)(\omega_o - \omega_k - i\gamma_o)}$$

$$+ \frac{1}{(\omega_o + \omega_i + i\gamma_o)(\omega_e + \omega_i + \omega_j + i\gamma_e)(\omega_o - \omega_k - i\gamma_o)}$$

$$\left. + \frac{1}{(\omega_o + \omega_i + i\gamma_o)(\omega_e + \omega_i + \omega_j + i\gamma_e)(\omega_o + \omega_\sigma + i\gamma_o)} \right]$$

$$+ \frac{Ne^4}{6\varepsilon_0\hbar^3} \langle\varphi_g|x|\varphi_o\rangle\langle\varphi_o|x|\varphi_e\rangle\langle\varphi_e|x|\varphi_{o2}\rangle\langle\varphi_{o2}|x|\varphi_g\rangle$$

$$\times \wp \left[ \frac{1}{(\omega_o - \omega_\sigma - i\gamma_o)(\omega_e - \omega_j - \omega_k - i\gamma_e)(\omega_{o2} - \omega_k - i\gamma_{o2})} \right.$$

$$+ \frac{1}{(\omega_o + \omega_i + i\gamma_o)(\omega_e - \omega_j - \omega_k - i\gamma_e)(\omega_{o2} - \omega_k - i\gamma_{o2})}$$

$$+ \frac{1}{(\omega_o + \omega_i + i\gamma_o)(\omega_e + \omega_i + \omega_j + i\gamma_e)(\omega_{o2} - \omega_k - i\gamma_{o2})}$$

$$+ \frac{1}{(\omega_o + \omega_i + i\gamma_o)(\omega_e + \omega_i + \omega_j + i\gamma_e)(\omega_{o2} + \omega_\sigma + i\gamma_{o2})}$$

$$+ \frac{1}{(\omega_{o2} - \omega_\sigma - i\gamma_{o2})(\omega_e - \omega_j - \omega_k - i\gamma_e)(\omega_o - \omega_k - i\gamma_o)}$$

$$+ \frac{1}{(\omega_{o2} + \omega_i + i\gamma_{o2})(\omega_e - \omega_j - \omega_k - i\gamma_e)(\omega_o - \omega_k - i\gamma_o)}$$

$$+ \frac{1}{(\omega_{o2} + \omega_i + i\gamma_{o2})(\omega_e + \omega_i + \omega_j + i\gamma_e)(\omega_o - \omega_k - i\gamma_o)}$$

$$\left. + \frac{1}{(\omega_{o2} + \omega_i + i\gamma_{o2})(\omega_e + \omega_i + \omega_j + i\gamma_e)(\omega_o + \omega_\sigma + i\gamma_o)} \right]$$

$$+\frac{Ne^4}{6\varepsilon_0\hbar^3}\langle\varphi_\text{g}|x|\varphi_\text{o2}\rangle\langle\varphi_\text{o2}|x|\varphi_\text{e}\rangle\langle\varphi_\text{e}|x|\varphi_\text{o2}\rangle\langle\varphi_\text{o2}|x|\varphi_\text{g}\rangle$$

$$\times \wp\left[\frac{1}{(\omega_\text{o2}-\omega_\sigma-i\gamma_\text{o2})(\omega_\text{e}-\omega_j-\omega_k-i\gamma_\text{e})(\omega_\text{o2}-\omega_k-i\gamma_\text{o2})}\right.$$

$$+\frac{1}{(\omega_\text{o2}+\omega_i+i\gamma_\text{o2})(\omega_\text{e}-\omega_j-\omega_k-i\gamma_\text{e})(\omega_\text{o2}-\omega_k-i\gamma_\text{o2})}$$

$$+\frac{1}{(\omega_\text{o2}+\omega_i+i\gamma_\text{o2})(\omega_\text{e}+\omega_i+\omega_j+i\gamma_\text{e})(\omega_\text{o2}-\omega_k-i\gamma_\text{o2})}$$

$$\left.+\frac{1}{(\omega_\text{o2}+\omega_i+i\gamma_\text{o2})(\omega_\text{e}+\omega_i+\omega_j+i\gamma_\text{e})(\omega_\text{o2}+\omega_\sigma+i\gamma_\text{o2})}\right] \quad (S6)$$

For the experimental condition of MIR pump-NIR reflection probe spectroscopy, the main terms for $\chi^{(3)}(-\omega;\Omega,-\Omega,\omega)$ are expressed as follows:

$$\chi^{(3)}_\text{main}(-\omega;\Omega,-\Omega,\omega)$$

$$=\frac{Ne^4}{6\varepsilon_0\hbar^3}\langle\varphi_\text{g}|x|\varphi_\text{o}\rangle\langle\varphi_\text{o}|x|\varphi_\text{e}\rangle\langle\varphi_\text{e}|x|\varphi_\text{o}\rangle\langle\varphi_\text{o}|x|\varphi_\text{g}\rangle$$

$$\times\left[\frac{1}{(\omega_\text{o}-i\gamma_\text{o}-\omega)^2(\omega_\text{e}-i\gamma_\text{e}+\Omega-\omega)}+\frac{1}{(\omega_\text{o}-i\gamma_\text{o}-\omega)^2(\omega_\text{e}-i\gamma_\text{e}-\Omega-\omega)}\right]$$

$$+\frac{Ne^4}{3\varepsilon_0\hbar^3}\langle\varphi_\text{g}|x|\varphi_\text{o}\rangle\langle\varphi_\text{o}|x|\varphi_\text{e}\rangle\langle\varphi_\text{e}|x|\varphi_\text{o2}\rangle\langle\varphi_\text{o2}|x|\varphi_\text{g}\rangle$$

$$\times\left[\frac{1}{(\omega_\text{o}-i\gamma_\text{o}-\omega)(\omega_\text{e}-i\gamma_\text{e}+\Omega-\omega)(\omega_\text{o2}-i\gamma_\text{o2}-\omega)}+\frac{1}{(\omega_\text{o}-i\gamma_\text{o}-\omega)(\omega_\text{e}-i\gamma_\text{e}-\Omega-\omega)(\omega_\text{o2}-i\gamma_\text{o2}-\omega)}\right]$$

$$+\frac{Ne^4}{6\varepsilon_0\hbar^3}\langle\varphi_\text{g}|x|\varphi_\text{o2}\rangle\langle\varphi_\text{o2}|x|\varphi_\text{e}\rangle\langle\varphi_\text{e}|x|\varphi_\text{o2}\rangle\langle\varphi_\text{o2}|x|\varphi_\text{g}\rangle$$

$$\times\left[\frac{1}{(\omega_\text{o2}-i\gamma_\text{o2}-\omega)^2(\omega_\text{e}-i\gamma_\text{e}+\Omega-\omega)}+\frac{1}{(\omega_\text{o2}-i\gamma_\text{o2}-\omega)^2(\omega_\text{e}-i\gamma_\text{e}-\Omega-\omega)}\right] \quad (S7)$$

The experimental result can be well-reproduced by this formula [red lines in Fig. 3(g,h)]. The calculation for $\chi^{(3)}(-\omega;\Omega,-\Omega,\omega)$ was actually conducted considering all terms, not just the main ones. Using Eq. (S7), $\Delta\tilde{\varepsilon}(\omega)$ was calculated from $\chi^{(3)}(-\omega;\Omega,-\Omega,\omega)$, and we can obtain $\tilde{\varepsilon}(\omega)+\Delta\tilde{\varepsilon}(\omega)$. From the static reflectivity spectrum $R(\omega)$ and the transient one $R(\omega)+\Delta R(\omega)$ calculated from $\tilde{\varepsilon}(\omega)+\Delta\tilde{\varepsilon}(\omega)$, the differential reflectivity spectrum $\Delta R/R\,(\omega)$ can be calculated [red line in Fig. 3(e,f)].